\documentclass[journal=jpcbfk,manuscript=article]{achemso}

\usepackage{chemformula} % Formula subscripts using \ch{}
\usepackage[T1]{fontenc} % Use modern font encodings
\usepackage{xspace}
\usepackage{mathtools}

\author{Jim Madge}
\affiliation{Department of Chemistry, Durham University, South Road, Durham DH1 3LE, United Kingdom}
\author{David Bourne}
\affiliation{Department of Mathematical Sciences, Durham University, South Road, Durham DH1 3LE, United Kingdom}
\author{Mark A.~Miller}
\affiliation{Department of Chemistry, Durham University, South Road, Durham DH1 3LE, United Kingdom}
\email{m.a.miller@durham.ac.uk}
\phone{+44 (0)191 3342037}

\title[Fragment competition in addressable assembly]
{Controlling Fragment Competition on Pathways to Addressable Self-Assembly}

\begin{document}

%%%%%%%%%%%%%%%%%%%%%%%%%%%%%%%%%%%%%%%%%%%%%%%%%%%%%%%%%%%%%%%%%%%%%
%% The "tocentry" environment can be used to create an entry for the
%% graphical table of contents. It is given here as some journals
%% require that it is printed as part of the abstract page. It will
%% be automatically moved as appropriate.
%%%%%%%%%%%%%%%%%%%%%%%%%%%%%%%%%%%%%%%%%%%%%%%%%%%%%%%%%%%%%%%%%%%%%
%\begin{tocentry}
%
%\includegraphics[width=90mm]{toc-fig-trimmed.pdf}
%
%\end{tocentry}

%%%%%%%%%%%%%%%%%%%%%%%%%%%%%%%%%%%%%%%%%%%%%%%%%%%%%%%%%%%%%%%%%%%%%
%% The abstract environment will automatically gobble the contents
%% if an abstract is not used by the target journal.
%%%%%%%%%%%%%%%%%%%%%%%%%%%%%%%%%%%%%%%%%%%%%%%%%%%%%%%%%%%%%%%%%%%%%
\begin{abstract}
Addressable self-assembly is the formation of a target structure from a set
of unique molecular or colloidal building-blocks, each of which occupies a
defined location in the target.  The requirement that each type of building-block
appears exactly once in each copy of the target introduces severe restrictions
on the combinations of particles and on the pathways that lead to successful
self-assembly.  These restrictions can limit the efficiency of
self-assembly and the final yield of the product.  In particular, partially
formed fragments may compete with each other if their compositions overlap,
since they cannot be combined.  Here, we introduce a ``completability''
algorithm to quantify competition between self-assembling fragments and use
it to deduce general principles for suppressing the effects of fragment
incompatibility
in the self-assembly of small addressable clusters.  Competition originates
from loops in the bonding network of the target structure, but loops may be
needed to provide structural rigidity and thermodynamic stability.
An optimal compromise can be achieved
by careful choice of bonding networks and by promoting semi-hierarchical
pathways that rule out competition between early fragments.  These concepts
are illustrated in simulations of self-assembly in two contrasting addressable
targets of 20 unique components each.
\end{abstract}

%%%%%%%%%%%%%%%%%%%%%%%%%%%%%%%%%%%%%%%%%%%%%%%%%%%%%%%%%%%%%%%%%%%%%
%% Start the main part of the manuscript here.
%%%%%%%%%%%%%%%%%%%%%%%%%%%%%%%%%%%%%%%%%%%%%%%%%%%%%%%%%%%%%%%%%%%%%

\newcommand{\nreplica}{\ensuremath{N_{\text{r}}} \xspace{}}
\newcommand{\nfragment}{\ensuremath{N_{\text{f}}} \xspace{}}
\newcommand{\ntype}{\ensuremath{N_{\text{t}}} \xspace{}}

\section{Introduction}

Self-assembly describes a broad class of processes, both naturally occurring and
human-made. In all cases, an ordered structure forms spontaneously and with high
fidelity from an initially disordered system of molecular or colloidal building-blocks.
Increasingly detailed control is now being exerted over target structures and
assembly pathways by the use of sophisticated building-blocks such as patchy
colloids\cite{pawar_fabrication_2010,yi_recent_2013,wang_colloids_2012}, and
DNA\cite{jones_programmable_2015}.
\par
DNA is a powerful starting-point for self-assembly because the binding
of complementary nucleotide sequences is so much stronger than that of sequences
that do not match.  Hence, different nucleotide sequences can be used to control
the pairwise interactions between building-blocks and thereby assemble structures from
multiple components.  This concept can be exploited by grafting DNA onto colloidal
particles\cite{alivisatos_organization_1996,mirkin_dna-based_1996,liu_self-organized_2016},
by folding long DNA strands like origami\cite{rothemund_folding_2006,hong_origami_review},
or by making tiles\cite{wei_complex_2012,Tikhomirov17a}
or bricks\cite{ke_three-dimensional_2012,ke_dna_2014,Ong17a} from short DNA helices.
These DNA-based approaches have introduced the paradigm of addressable
self-assembly\cite{frenkel_order_2015}, in which each component of a complex structure
is unique and has a specified location.
\par
Addressable assembly has the advantage of providing site-by-site control over the
properties of the final structure.  It also avoids the need to rely on symmetry, since
building-blocks that are unique may each have environments that are also unique, making it
possible to assemble structures of arbitrary complexity.  In this sense, addressable
assembly lies at the opposite extreme from some well-known examples of self-assembly
such as virus capsids\cite{madge_design_2015}.  In capsid assembly, the high
symmetry of the protein shell (often icosahedral) provides a large number of
equivalent or quasi-equivalent molecular environments, so that the shell can form from
multiple copies of just one building-block in some
cases\cite{johnson_quasi-equivalent_1997}.
\par
In a fully addressable structure, each type of building-block appears only once, and
this constraint introduces some important new entropic
considerations\cite{frenkel_order_2015,cademartiri_programmable_2015}.
Self-assembly of a target structure always requires a specific number $N$ of
building-blocks to come together.  In a one-component system, any $N$ available monomers
are acceptable and they may appear in the structure in any order.  In
contrast, addressable assembly requires a specific combination of building-blocks
(one of each type) and each has a particular location within the target.  Hence,
the number of acceptable particle permutations in the fully assembled system is greatly
reduced in the addressable case, lowering the statistical weight of the assembled state.
Likewise, the number
and nature of pathways leading to correct assembly are dramatically altered.  Two
consequences of these characteristics are that addressable assembly is usually only
viable within a narrow temperature window, and that the yield may be
low\cite{Ong17a,madge_design_2015,reinhardt_numerical_2014,Sajfutdinow18a}.
\par
Theoretical and computational
work\cite{reinhardt_numerical_2014,jacobs_rational_2015,jacobs_communication:_2015,jacobs_self-assembly_2016}
has shown that addressable assembly can be understood as a non-classical nucleation
process.  The existence of a nucleation free-energy barrier and a critical
cluster assists correct assembly partly by slowing down the approach to equilibrium
and by providing the opportunity for incorrectly formed fragments to break up.
A time-dependent temperature protocol can help assembly, starting with nucleation
at a higher temperature and proceeding with growth at a lower
temperature\cite{jacobs_rational_2015,Fullerton16a,Sajfutdinow18a},
even if the cooling schedule spans a very narrow range\cite{Ong17a}.
\par
A particular feature of addressable assembly is that correct fragments of the target
can begin to form but then compete with each other because they have overlapping
compositions that prevent them from combining or from all reaching completion.  Assembly
in a one-component system is less prone to this problem because any monomer may
contribute to growth of a structure\cite{madge_design_2015}.  This problem can be
partly mitigated by the nucleation kinetics described above.  Nevertheless, it should
be possible to raise yields by positively avoiding competition between partially formed fragments.
\par
With some notable exceptions\cite{Wales17a,Fejer2018,Halverson13a,Halverson17a}, most
computational work on addressable assembly has concentrated on the formation
of just one copy of the target structure, starting from
the perfect stoichiometry of only one building-block of each type.  Hence, such studies do not
explicitly account for fragment competition.  Here, we focus directly on this intrinsic
aspect of addressable assembly.  We introduce a metric for quantifying fragment competition in
terms of a ``completability index'' that describes the instantaneous state of the system.
This index helps to identify where competition is the limiting factor in self-assembly.
We then show how competition can be reduced by certain choices for the topology and strength
of the bonding network in the target structure.  General principles emerge that
allow higher yields to be reached, and that make assembly robust over a wider range of
temperatures.
\par
The rest of the article is structured as follows.  We start by
specifying a variant of an idealized coarse-grained model and a dynamics-like Monte Carlo (MC)
method\cite{madge_design_2015,madge_minimal_2017} for studying addressable assembly.
We then show how to quantify fragment competition by introducing the completability index and
algorithm.  We use these tools to analyze the self-assembly
of two contrasting target structures.  Finally, we summarize
the general principles and conclusions that can be extracted from the study.

\section{Model}
\label{sec:str-model}

\subsection{Patchy potential}
\label{sec:str-potential}
In previous work, we introduced a generic off-lattice model to compare design
strategies for self-assembly\cite{madge_design_2015} and to optimize the
design of building-blocks with controllable complexity\cite{madge_minimal_2017}.
The building-blocks were hard cubes whose faces were patterned with attractive
patches that drive the self-assembly.
Here, we use a variant of the patchy cube model to focus on fragment competition.
The key modifications are simplifying the patterned faces to a single interaction
site per face, and placing the minimum of the pairwise attraction slightly away
from contact between the hard cores to avoid artificially severe steric requirements
in closely packed targets.  Our model has much in common with the off-lattice version
of a model studied by Jacobs, Oxtoby and Frenkel\cite{jacobs_phase_2014}.
\par
The patchy cubes have edges of length $d$.  The cores are impenetrable and overlap
detection is handled in the simulations by treating them as oriented bounding
boxes\cite{gottschalk_obbtree:_1996,smallenburg_vacancy-stabilized_2012,john_phase_2005}.
Each face may have up to one attractive site.  Pairwise interactions between these
patches on different cubes are attenuated by both an angular and a torsional factor
such that the minimum occurs when the interacting faces are parallel and the cubes
have a particular mutual orientation with respect to the line that joins their centers.
The angular attenuation causes each patch to be associated with a particular face
of the building-block and captures the resulting directionality of the attractive
interactions.  The torsional contribution to the potential accounts for the overall
effect of a more detailed description of the interactions\cite{wilber_monodisperse_2009}
in terms of a pattern of interactions on each face of the cubes.
Such patterns were previously represented explicitly when
they were the main focus of an investigation\cite{madge_design_2015,madge_minimal_2017}.
\par
The pairwise interaction is an attractive well that
switches between a Morse potential and Gaussian repulsion at the minimum
of the potential.  The value, gradient and curvature of the two parts are
matched at the point of hand-over.  The functional form of the interaction is
\begin{equation}
  V_{ij}^{\text{M}}(r_{ij}) =
  \begin{cases}
    \varepsilon_{ij}
    \left[
      e^{-2\alpha(r_{ij} - 5d/4)} - 2 e^{-\alpha(r_{ij} - 5d/4)} \right] 
    & \text{if} \, r_{ij} > 5d/4
    \\
    \varepsilon_{ij} \exp\left( -\alpha^{2} [r_{ij} -5d/4]^{2} \right) 
    & \text{if} \, d < r_{ij} \le 5d/4
  \end{cases}
  \label{eq:morse}
\end{equation}
where $r_{ij}$ is the distance between patches $i$ and $j$.  $\alpha$ is a
parameter that controls the range of the potential, which we fix\cite{madge_design_2015}
at $\alpha=6d^{-1}$, which means that the curvature of the potential at its
minimum is the same as that of the Lennard-Jones potential.  $\varepsilon_{ij}$
determines the depth of the potential at its minimum.  For the addressable systems
considered in this work, the identity of the building-blocks is effectively
specified by the set of $\varepsilon_{ij}$ values for all combinations of patches.
Non-zero values of this parameter determine which faces of the different particles
bind to each other and how strongly they do so.
\par
\begin{figure}[t]
  \centering
  \includegraphics[width=75mm]{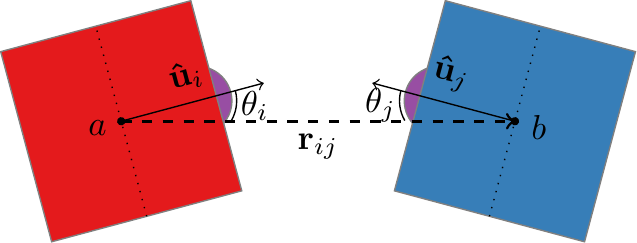}
  \caption{
    Schematic representation of the interaction between patches $i$ and $j$ on two cubes
    $a$ and $b$, showing the definition of the angles $\theta_i$ and $\theta_j$.
  }
  \label{fig:patches}
\end{figure}

The interaction site for
each patch is embedded in the particle perpendicular to the face on which the
patch sits at a depth of $d/2$ (Fig.~\ref{fig:patches}).  The distance dependence
of the exponent in Eq.~(\ref{eq:morse}) therefore ensures that the minimum of the
interaction between two patches occurs when two particles faces are separated by
a distance of $d/4$.  This separation helps ensure that the hard repulsion of
particles does not artificially prevent dense clusters from forming.
The potential is truncated at a distance of $2d$. To avoid a discontinuity at
the cut-off, the potential is shifted by $V_{ij}^{\text{M}}(2d)$. The potential
is then rescaled so that the value of the potential at the minimum is
$\varepsilon_{ij}$.
\par
The angular attenuation of the potential is a Gaussian of the form
\begin{equation}
  V^{\text{ang}} (\hat{\bf r}_{ij}, \hat{\bf u}_{i}, \hat{\bf u}_{j}) = 
  \exp \left( - \frac{\theta_{i}^{2} + \theta_{j}^{2}}{2 \sigma_{\text{ang}}^{2}}
  \right) \, ,
  \label{eq:ang}
\end{equation}
where $\hat{\bf r}_{ij}$ is the unit vector from patch $i$ to patch $j$
(Fig.~\ref{fig:patches}).  $\theta_i=\cos^{-1}(\hat{\bf r}_{ij}\cdot\hat{\bf u}_i)$ and
$\theta_j=\cos^{-1}(\hat{\bf r}_{ji}\cdot\hat{\bf u}_j)$ are the angles between
$\hat{\bf r}_{ij}$ and patches $i$ and $j$, respectively.  $\hat{\bf u}_{i}$ and
$\hat{\bf u}_{j}$ are the vectors normal to the faces upon which patches $i$ and
$j$ sit (Fig.~\ref{fig:patches}).  $\sigma_{\text{ang}}$ determines how quickly
the potential decays with any deviation from perfect alignment and is
set at $0.2$ in this work\cite{madge_design_2015}.
The embedded interaction sites with
angular attenuation are similar to a previous model for patchy spherical
particles\cite{doye_controlling_2007,villar_self-assembly_2009,wilber_self-assembly_2009}.
\par
Each patch is also given an intrinsic orientation, pointing from the center of
the face to which the patch belongs to the center of one of the face's edges.
The torsional attenuation of the interaction between two patches depends on the
torsional angle between their orientation vectors, with the Gaussian form
\begin{equation}
  V^{\text{tor}} ( {\bf r}_{ij}, \hat{\bf v}_{i}, \hat{\bf v}_{j} ) = 
  \exp \left(
  \frac{
    - \varphi_{ij}^{2}
  }{
    2 \sigma_{\text{tor}}^{2}
  }
  \right)
  \, ,
  \label{eq:tor}
\end{equation}
where $\varphi_{ij}$ is the torsional angle between the two orientation vectors
(${\hat{\bf v}}_{i}$ and ${\hat{\bf v}}_{j}$) of patches $i$ and $j$.  The width
of the Gaussian, $\sigma_{\text{tor}}$, controls how quickly the potential is
attenuated as the torsional angle rotates from $0$, and is set here
to $0.7$.
\par
The overall form of the potential is therefore
\begin{equation}
  V^{\rm patch}_{ij}({\bf r}_{ij},\hat{\bf u}_i,\hat{\bf u}_j) = 
  \left[
    \frac{
      V^{\rm M}_{ij}(r_{ij}) - V^{\rm M}_{ij}(2d)
    }{
    \varepsilon_{ij} - V^{\rm M}_{ij}(2d)
    }
  \right]
  \Theta(2d-r_{ij})
  V^{\text{ang}}(\hat{\bf r}_{ij},\hat{\bf u}_i,\hat{\bf u}_j) 
  V^{\text{tor}}({\bf r}_{ij},\hat{\bf v}_{i},\hat{\bf v}_{j})
  \, ,
  \label{eq:potential}
\end{equation}
where ${\bf r}_{ij} = r_{ij} \hat{\bf r}_{ij}$ and $\Theta$ is the Heaviside step
function, which enforces the cut-off at $r_{ij} = 2d$.  The total interaction
energy between any two particles is given by
\begin{equation}
V^{\rm cube}_{ab}({\bf r}_{ab}, {\boldsymbol\Omega}_a, {\boldsymbol\Omega}_b)
= \sum_{i\in a} \sum_{j\in b} V^{\rm patch}_{ij}({\bf r}_{ij},\hat{\bf u}_i,\hat{\bf u}_j)
\Delta_{ij}(\hat{\bf r}_{ij},{\boldsymbol\Omega}_a,{\boldsymbol\Omega}_b) \, ,
\end{equation}
where $\boldsymbol{\Omega}_a$ represents the orientation of particle $a$ and
${\bf r}_{ab}$ is the vector from the center of particle $a$ to the center of
particle $b$.  $\Delta_{ij} = 1$ if the faces on which patches $i$ and $j$ sit
are the closest pair of most aligned faces of the two cubes, and $0$ otherwise.
$\Delta_{ij}$ therefore acts as an angular truncation of the potential.
The strength of the interaction is negligible at the point of truncation
(typically less than $10^{-6}\varepsilon_{ij}$).

\subsection{Dynamical Monte Carlo algorithm}
\label{sec:str-monte}

Dynamical simulations of the patchy cube model are performed using our
hybrid MC algorithm, which combines bulk diffusion moves with internal
relaxation moves to produce trajectories with correct relative diffusion rates for
clusters of different sizes as well as collective internal motion.
The algorithm is specifically designed to cope with the inhomogeneous structure
of a self-assembling system and has been described in full previously\cite{madge_design_2015}.
Here we summarize its key features.
\par 
Diffusion moves act on entire isolated clusters, defined as collections of particles
connected by non-zero energetic interactions.  One such cluster
is uniformly chosen to be moved.  Half of the diffusion moves are
translational, and the other half rotational.
A translational diffusion move is constructed by picking a random displacement
on a Gaussian distribution in each dimension.  For a rotational move, a similar
approach is used to generate random rotation vectors about the
cluster's center of mass\cite{romano_monte_2011}.
The magnitudes of translations and rotations are scaled
to account for two factors.  First, under Brownian motion, the translational and
rotational diffusion constants of spherical particles depend on the radius of
the sphere.  For this purpose, our clusters are approximated as spheres with radius
proportional to $n^{\sfrac{1}{3}}$ where $n$ is the number of particles in the
cluster.  Second, the size of moves is scaled to ensure that the quotient of
translational to rotational diffusion respects the Stokes--Einstein relations.
\par 
The internal relaxation of clusters is handled with the symmetrized Virtual Move
Monte Carlo (VMMC)
algorithm\cite{whitelam_avoiding_2007,whitelam_role_2009}.
VMMC begins by picking a seed particle and then recruits neighboring
particles for the proposed translation or rotation according to the change in
potential energy incurred.  This approach therefore effectively accounts for
forces through the approximate gradient of the potential.
\par 
In all simulations we will use a reduced temperature $T^*$, defined as
\begin{equation}
  T^* = k_{\rm B} T / \varepsilon
  \, ,
  \label{eq:Tstar}
\end{equation}
where $\varepsilon$ is the mean minimum interaction energy between all pairs of
patches in the target structure, $\left< \varepsilon_{ij} \right>_{i,j}$.
\par
The reduced temperature is also used to define a relative time scale for our
simulations.  The dynamical algorithm described in this section provides control
over the diffusion rates of aggregates in the simulation.  The rate of diffusion is
determined by the width of the distribution from which moves are
selected, which we fix at $0.2d$, giving a good
acceptance of the internal relaxation moves across a wide range of temperatures.
However, Stokes--Einstein diffusion constants are proportional to temperature, and
so we must consider the reduced temperature when defining the reduced time.  The
reduced relative time is taken as
\begin{equation}
  t^* = s / T^*
  \, ,
  \label{eq:tstar}
\end{equation}
where $s$ is the number of MC sweeps completed and a sweep in an
$N$-particle system consists of $N$ MC trial moves.

\section{Quantifying Fragment Competition}
\label{sec:str-complete}

\subsection{Completability Index}

In order to assess the state of a system in the process of assembling into
multiple copies of a target structure, we introduce a completability index,
which identifies competition between fragments that may prevent a system from reaching
maximum yield.  The analysis can be
applied to any instance of addressable assembly, {\em i.e.}, where each particle in
the target structure is unique and so appears exactly once per copy. 
\par
\begin{figure}[t]
  \centering
  \includegraphics[width=75mm]{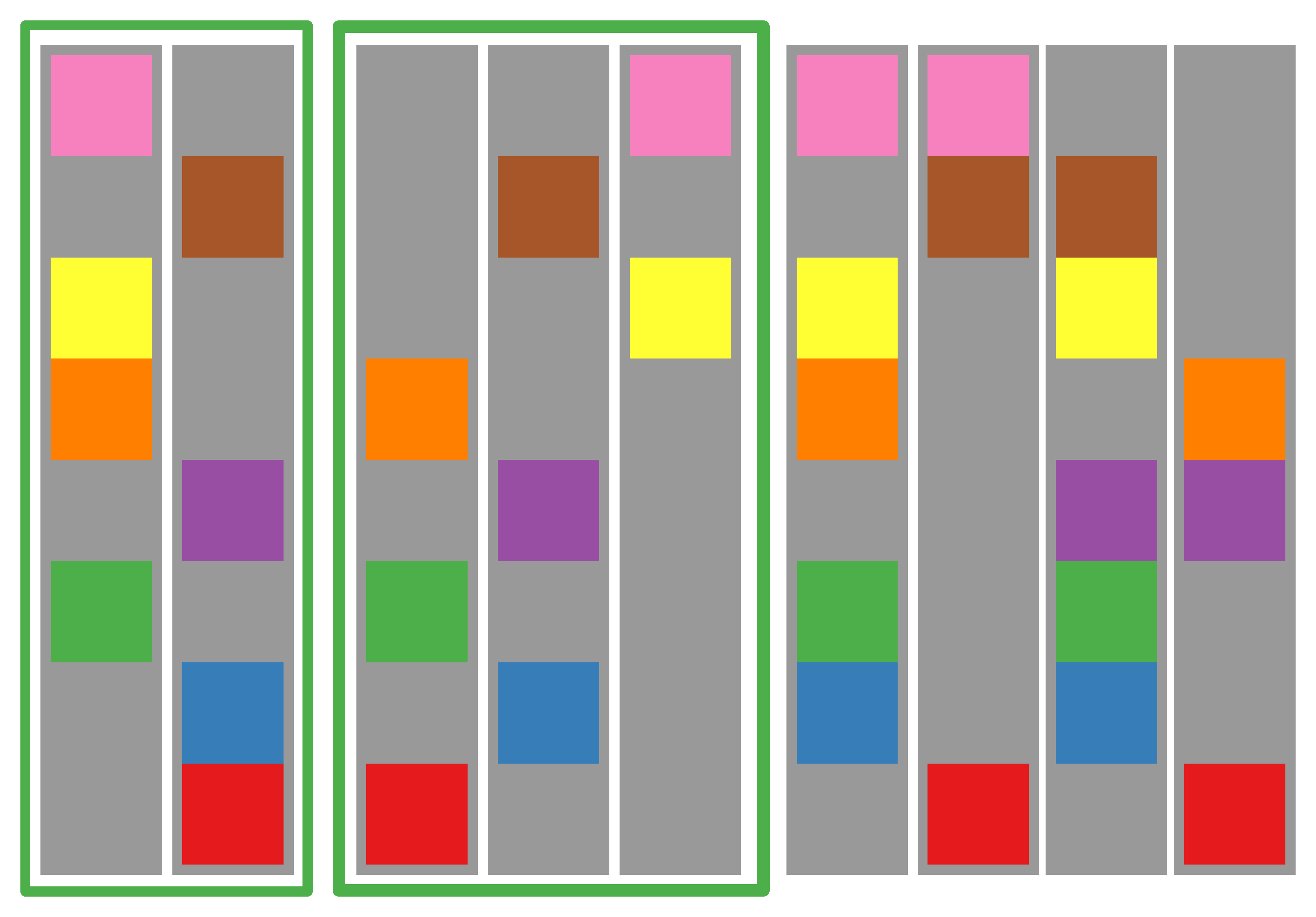}
  \caption{
    An example of the completability algorithm for a cluster of
    eight unique components, identified here by color.  There are four copies of
    each component, enough to make a maximum of four complete clusters.  At a given
    instant the building-blocks may have assembled into multiple fragments,
    indicated in this example by vertical gray bars.
    The job of the completability algorithm is to determine the maximum number of
    complete targets (involving one component of each color) that can be formed by
    the fragments, which in this case is two, as indicated by the green boxes.
    No further targets or larger aggregates
    may be made from the remaining fragments because they have overlapping compositions.
  }
  \label{fig:complete}
\end{figure}
The algorithm identifies all aggregates present in a given configuration (defined
by non-zero energetic interactions between the building-blocks), and
determines the maximum number of complete target clusters that can be formed by
combining them.  An example of the task is shown in Fig.~\ref{fig:complete}, where there
are enough particles to build four target structures, but the existing fragments can
only be combined to complete two of them.  The task is a combinatorics problem that can
be expressed in the formalism of linear programming (LP).
An LP problem is a constrained optimization problem where the objective function is linear
and the constraints are linear equalities or inequalities.
LP problems take the general form
\begin{equation}
  \begin{split}
    \text{maximize} & \quad {\bf c}^{\rm T} {\bf x} \\
    \text{subject to} & \quad {\bf A}{\bf x} \le {\bf b} \\
    \text{and} & \quad {\bf x} \ge {\bf 0}
    \, .
  \end{split}
  \label{eq:lp}
\end{equation}
Here, ${\bf x}$ is a variable vector of length $m$ whose components are to be
optimized in order to find the constrained maximum value of
${\bf c}^{\rm T}{\bf x}$. The
vectors ${\bf c}$ and ${\bf b}$ are of length $m$ and $n$ respectively and their
values are both fixed.  ${\bf A}$ is a given
$n\times m$ matrix.  The vector ${\bf c}$ contains the coefficients of the
variables in the objective function (${\bf c}^{\rm T} {\bf x}$), while
${\bf A}$ and ${\bf b}$ embody the constraints.
\par
LP is commonly used to solve assignment problems, which typically take the
form of assigning $N$ workers to $N$ tasks in the optimal way, given that the
performance of different workers on different tasks may vary.  In the context of
addressable assembly, we express the sorting of fragments from the output of a
simulation into the largest possible number of target clusters as a binary LP
problem (where the components of the vector ${\bf x}$ are all either 1 or 0).
\par
To analyze the completability of a given snapshot from a simulation,
we first need to obtain a list of all fragments that it contains,
along with their compositions.  We must also test whether each of these
fragments is valid.  A valid cluster is a correctly structured sub-fragment
of the target cluster\cite{madge_design_2015}, one requirement of which is
that it contains no more than one of each particle type.
Any fragments that are not valid must be excluded from the analysis, and the
number of possible targets reduced accordingly.  For example, if the target
cluster consists of four particles {\it ABCD}, and there are ten copies of each particle
in the simulation, but we have identified and discounted the invalid cluster
{\it ABA}, then it is now only possible to complete a maximum of eight targets,
since two {\it A} particles have been removed.
\par
Given the set of identified fragments, we solve the following problem.
We define a number of sets, equal to the maximum number of possible targets.
Each fragment may be placed in a set,
taking with it all of its particles.  We define the score of a set as the
number of particles it contains, and an overall objective function ${\cal N}$
as the sum of all set scores, {\em i.e.}, the total number of
particles placed into sets.  We then maximize ${\cal N}$ by varying the
assignment of fragments to sets, subject to the constraints that each fragment
may only appear in at most one set, and that each set may only contain at most
one of each type of particle.
Finally, we define the completability index of the fragments as the ratio of the
number of completable sets to the maximum number of targets that could, in
principle, be built from the original monomers.  In the example of
Fig.~\ref{fig:complete} only two complete targets can be obtained from the
fragments, but there are four copies of each monomer, so the completability
is $\frac{1}{2}$.

\subsection{Algorithm}

Here, we show how to formulate the fragment assignment as an LP problem like that in
Eq.~(\ref{eq:lp}).
To omit the formal presentation of the completability algorithm, readers may skip
to the Results section at this point.
\par
We start by expressing the fragments in a given configuration
as a binary string.  For instance, in an eight-component cluster with building-blocks
{\it A}--{\it H}, the fragment {\it ABDF} would be $(1,1,0,1,0,1,0,0)$. The sum of the
string gives an indication of how complete the fragment is, with only a correct
target achieving the maximum value.
\par
Now consider a system of \nreplica copies of each \ntype types of building-blocks,
divided into \nfragment independent fragments.  Note that \nreplica is also the
maximum number of targets that can be created simultaneously.  Similarly to the
fragment strings, the assignment of fragments to sets (combinations of fragments
in our proposed solution) may be expressed as a binary vector of
length \nfragment.  For example, in a system with four fragments, the set
vector $(0,1,0,1)$ would imply that fragments $2$ and $4$ belong to the set,
but that $1$ and $3$ do not.
\par
The assignment of the \nfragment fragments into sets may therefore be
represented by a concatenation of all the \nreplica individual set vectors.
There are \nreplica set vectors, as this is the maximum number of possible targets.
The assignment vector ${\bf x}$ therefore has the form
\begin{equation}
  {\bf x} = 
  \begin{pmatrix}
\begin{rcases*}
%   \left.
    \begin{pmatrix}
      \vdots
    \end{pmatrix}
%   \right\} & {\bf s}_{1}
\end{rcases*}
{\bf s}_{1}
   \\
\begin{rcases*}
%   \left.
   \begin{pmatrix}
     \vdots
   \end{pmatrix}
%   \right\} & {\bf s}_{2}
\end{rcases*}
{\bf s}_{2}
   \\
   \vdots &
   \\
\begin{rcases*}
%   \left.
   \begin{pmatrix}
     \vdots
   \end{pmatrix}
%   \right\} & {\bf s}_{\nreplica}
\end{rcases*}
{\bf s}_{\nreplica}
  \end{pmatrix}
  \, ,
  \label{eq:vec-assign}
\end{equation}
where ${\bf s}_{1}$ is the first set vector. ${\bf x}$ has length $\nfragment \nreplica$.
\par
We can collect the scores (sum of the binary strings) of the \nfragment
fragments in a vector,
\begin{equation}
  {\bf n} = 
  \begin{pmatrix}
    n_{1} \\
    n_{2} \\
    \vdots \\
    n_{\nfragment}
  \end{pmatrix}
  \, ,
  \label{eq:nvec}
\end{equation}
where $n_{1}$ is the score of the first fragment.
We may also define the vector ${\bf c}$, as \nreplica repeats of ${\bf n}$,
\begin{equation}
  {\bf c} = 
  \begin{pmatrix}
    {\bf n} \\
    {\bf n} \\
    \vdots \\
    {\bf n}
  \end{pmatrix}
  \, ,
  \label{eq:frag-vec}
\end{equation}
which has the same length, $\nfragment \nreplica$, as ${\bf x}$.
The column vectors ${\bf c}$ and ${\bf x}$ define our objective function,
$\cal{N}={\bf c}^{\rm T} {\bf x}$,
which is equal to the total number of particles
placed into sets.  The maximum of this function corresponds to all particles
being placed into sets, meaning that all $\nreplica$ target structures
are complete.  We therefore want to maximize $\cal{N}$
subject to two constraints:
\begin{enumerate}
  \item No fragment appears more than once in ${\bf x}$.
  \item No type of building-block appears more than once in any set.
\end{enumerate}
\par
To construct the constraint matrix ${\bf A}$, we first introduce the
matrix $\tilde{\bf A}$ with dimensions $\ntype \times \nfragment$ to encode
the compositions of the fragments.  Element
$\tilde{A}_{ij}$ of $\tilde{\bf A}$ is $1$ if fragment $j$ contains
building-block type $i$, and $0$ if not.  The constraint matrix is then given
in block form by $\nreplica$ copies of $\tilde{\bf A}$ and $\nreplica$ copies
of the $\nfragment\times\nfragment$ identity matrix ${\bf I}$ in the
arrangement,
\begin{equation}
  {\bf A} = 
  \begin{pmatrix}
    \tilde{\bf A} & & & \\
    & \tilde{\bf A} & & \\
    & & \ddots & \\
    & & & \tilde{\bf A} \\
    {\bf I} & {\bf I} & \cdots & {\bf I}
  \end{pmatrix}
  \, .
  \label{eq:Amatrix}
\end{equation}
${\bf A}$ therefore has dimensions
$(\nreplica\ntype + \nfragment) \times \nfragment\nreplica$
and the product ${\bf A}{\bf x}$ has $\nreplica\ntype + \nfragment$
elements.  The first $\nreplica\ntype$ elements of ${\bf A}{\bf x}$
give the number of building-blocks of each type in each set, starting
with the number of {\it A} blocks in set 1, the number of {\it B} blocks
in set 1, {\it etc.}, then moving onto the number of each type in set 2,
{\it etc.} The remaining \nfragment elements of ${\bf A}{\bf x}$
give the number of times
each fragment has been assigned to any set, in the order fragment 1,
fragment 2, {\it etc.}
\par
The two constraints are fulfilled so long as every element of ${\bf A}{\bf x}$
is less than or equal to 1.  Considering the first inequality in Eq.~(\ref{eq:lp}),
it follows that the constraint vector ${\bf b}$ is simply a vector of length
$(\nreplica\ntype + \nfragment)$ where every element is equal to 1.  We now
have the definitions of the vectors and matrices required by
Eq.~(\ref{eq:lp}) (${\bf c}$, ${\bf x}$, ${\bf b}$ and ${\bf A}$) in order to optimize
$\cal{N}={\bf c}^{\rm T} {\bf x}$.
Maximization of this function will sort
fragments into sets in such a way as to complete as many copies of the target structure
as possible.  The number of complete targets can then be determined by inspecting
the elements of ${\bf x}$ after optimization.
\par
If an invalid fragment arises,
then the total number of targets that can be formed is now less
than the number of replicas of each building-block type.  In such a case, using
the formalism presented here, the value of \nreplica must be reduced
appropriately, which in turn reduces the number of sets to be created in the
vector ${\bf x}$.
\par
The global maximum of the objective function $\cal{N}$ may not be unique.  If
$\cal{N}$ takes its largest possible value of $\nreplica\ntype$ then any multiple
solutions simply correspond to different ways of combining the fragments to
create complete targets.  However, when it is not possible to resolve the
system into a full contingent of completed targets, a set of degenerate maxima
may exist.  While some maxima may correspond to the largest number of completed
targets, others may correspond to a larger number of incomplete sets, such that
the same number of particles placed into sets overall is the same.  To
ensure that the maximum number of targets has been found, we use
the following procedure.  If a full contingent of targets is not found then an
attempt is made to build one more than was found previously.  This is done by
reducing the size of the matrices to include a smaller number of fragment sets.
If a solution exists with this reduced number of complete targets, then the algorithm is
forced to find it because any incomplete sets would decrease the overall score.
The number of targets is then incremented by one again, and the test is repeated.
When it is not possible to build the incremented number of targets, the number
actually found is the true maximum.  Therefore this guarantees that the maximum
number of targets is found, provided that the algorithm used to solve the binary
LP problem does indeed return a global maximum.
\par
A detailed example of the completability algorithm, including the explicit form
of the matrix equations, is given in the Supporting Information as an
illustrative case.

% The formulation of this assignment as an
% LP problem like that in Eq.~(\ref{eq:lp}) is presented in
% the Appendix.  The maximum number of target structures that can be
% obtained by combining the available fragments is the number of complete sets
% (those containing one particle of every type) in the optimized assignment.
% We define the completability index of the fragments as the ratio of the
% number of completable sets to the maximum number of targets that could, in
% principle, be built from the original monomers.  In the example of
% Fig.~\ref{fig:complete} only two complete targets can be obtained from the
% fragments, but there are four copies of each monomer, so the completability
% is $\frac{1}{2}$.

\section{Results}
\label{sec:str-result}
\begin{figure}[t]
  \centering
  \includegraphics[width=75mm]{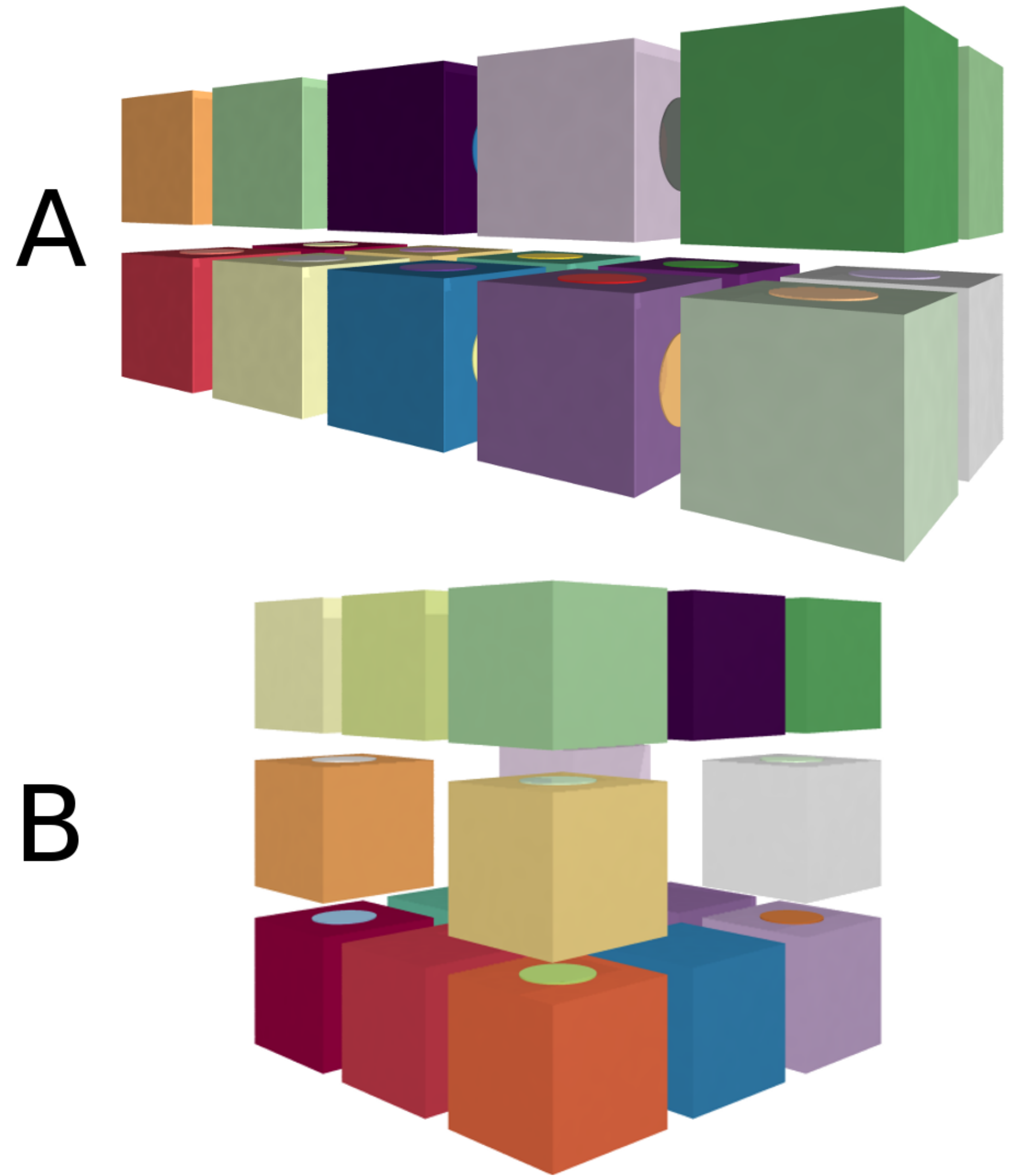}
  \caption{
    The two target clusters used in this work, both consisting of twenty
    unique particles: a compact structure A (top), and an open cage structure
    B (bottom).
  }
  \label{fig:clusters}
\end{figure}
We will use two test clusters to investigate the impact of target geometry,
connectivity and bond-strength heterogeneity on self-assembly.
The idealized (lowest-energy) geometry of the clusters is shown in
Fig.~\ref{fig:clusters}.
Both clusters are comprised of twenty particles, each of which is unique.
Target A (Fig.~\ref{fig:clusters} top) is a compact $5\times2\times2$ cuboid of
particles.  In contrast, target B (Fig.~\ref{fig:clusters} bottom) has an open,
cage-like structure, resembling a $3\times3\times3$ cube with the face- and
body-center particles absent.  In the first instance, the adjacent faces of
all neighboring particles interact equally and specifically, {\em i.e.},
$\varepsilon_{ij}=\varepsilon$ for faces that are in contact in Fig.~\ref{fig:clusters}
and $\varepsilon_{ij}=0$ otherwise, thereby encoding addressability into the
building-blocks.
\par
Each self-assembly data-point presented here is the mean of 25 dynamical MC
simulations at each set of conditions.  Each simulation begins from a distinct
starting configuration consisting of a disordered fluid of monomers, generated
from a single high-temperature simulation.  Self-assembly is initiated
by an instantaneous quench to the assembly temperature of interest.
In all simulations, each component is present at a number density of $0.002d^{-3}$.
All the simulations contain $80$ particles, enough to make a total of $4$
targets.  Simulating multiple replicas of each component is essential
to observe fragment competition, which many simulation studies of addressable
assembly have neglected.  Even so, we note that any study of self-assembly in
a finite system with a fixed number of particles cannot capture all relevant
density fluctuations\cite{Ouldridge10b}.  The standardized conditions of the
simulations described here facilitate comparisons within this limitation.

\subsection{Target clusters}
\label{sec:str-res-clusters}
\begin{figure}[t]
  \centering
  \includegraphics[width=75mm]{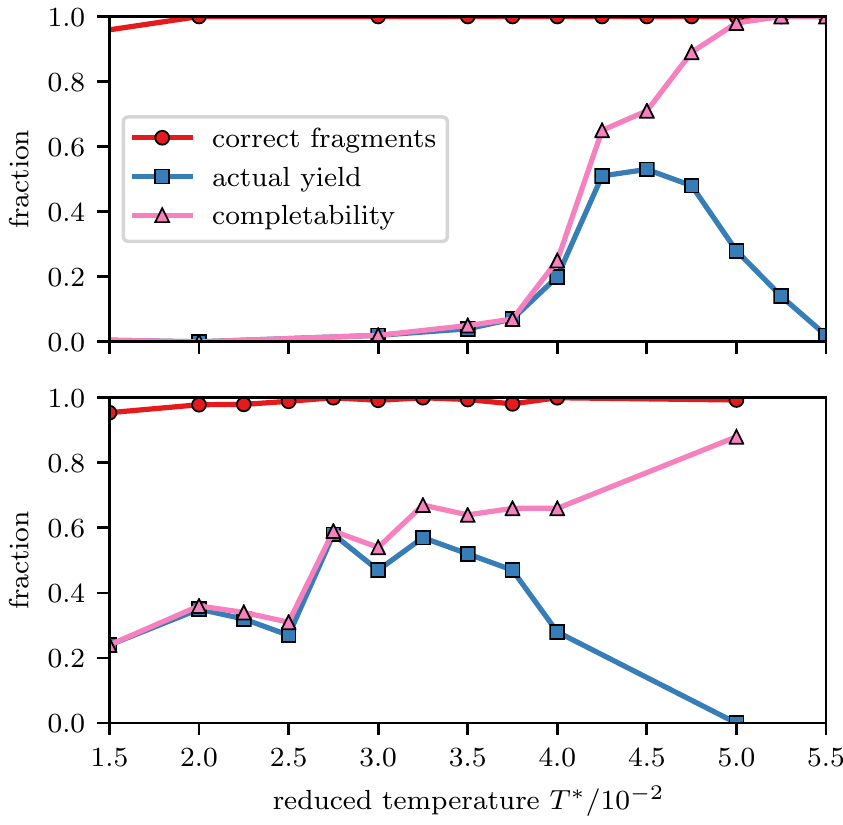}
  \caption{
    The yields of complete targets and correct partial fragments, and the
    completability index for target A (top) and target B (bottom).
  }
  \label{fig:frags_clusters}
\end{figure}
The yields of both target clusters after an assembly time of $t^* =1.33\times10^8$
are shown as a function of the assembly temperature in Fig.~\ref{fig:frags_clusters}.
This time gives the building-blocks a chance to respond to the assembly temperature and
allows us to monitor kinetic progress even though assembly may continue beyond this
point of observation.
In these plots, the mean yield, which is the fraction of particles in complete and
correct clusters, is shown by the blue squares. 
The pink triangles show the mean completability index of the fragments present
at the end of the simulation, and indicate the frustration present in
the system due to fragments with mutually incompatible compositions.
Such fragments must break up before their building-blocks can be combined into
complete targets.
The red circles indicate the fraction of particles in correct fragments, {\em i.e.},
complete targets and sub-fragments of that target in which all particles are in the
correct positions and mutual orientations.
\par
For target A (Fig.~\ref{fig:frags_clusters} top) optimal assembly occurs within a
narrow temperature window, $T^* \approx [4.25, 4.5]\times10^{-2}$.  Above this
temperature range the cluster is thermodynamically unstable, and the system exists
mainly as small aggregates and monomers.  This is discernible from the combination of
low actual yield but high completability in Fig.~\ref{fig:frags_clusters}.
At temperatures below optimal, this cluster suffers from severe frustration, which
limits the extent of assembly possible.  The close pairing of the yield and
completability lines shows that assembly has proceeded as far as possible with
the existing fragments, and that some fragments must first break up for further
progress to be made.  Incomplete fragments become increasingly stable at low
temperatures, making frustration harder to correct and effectively capping the
maximum yield at very low values.
Yields at optimal temperatures are modest at approximately $50\%$, and lie below
the maximum completability.  This implies that assembly has become slow for this structure,
and that there may be steric effects that stall assembly as fragments near completion. 
\par
The equivalent plot for target B has a different character, showing a completability
index that decays much more gradually with decreasing temperature.
Yields in the optimal range of $T^* \approx [2.75, 3.75]\times10^{-2}$
are similar to those for target A.  For both targets, the fraction of particles in
correct fragments is always high, showing that self-assembly is not inhibited by
erroneous binding or uncontrolled growth of fragments.

\subsection{Bonding topology}
\label{sec:str-res-topology}
These preliminary results show that frustration between incompatible fragments
is a significant factor in limiting the self-assembly of these target clusters.
We will now explore the origin of the problem and demonstrate that it is strongly
affected by the characteristics of the bonding network.
\par
The sparsest possible bond network that still connects the cluster is a linear
chain that may need to be branched but contains no loops.
The advantage of such arrangements is that they make it
impossible for the system to aggregate into incompatible clusters, provided
that there are equal numbers of each component.  However, the addition of just
one more link introduces a loop and allows frustrated clusters to arise.
For example, consider a target composed of four components {\it A}--{\it D} arranged in a
square, and a system containing eight particles, enough to form two complete
targets.  The particles might be arranged into the aggregates {\it ABC}, {\it AB}, {\it CD}, {\it D}.
In the case where the cluster is linearly connected, the only bonding interactions
are {\it A}--{\it B}, {\it B}--{\it C} and {\it C}--{\it D}.
We may combine clusters {\it ABC} with {\it D} and {\it AB} with {\it CD} to form two complete
targets.  This is the only way to combine these fragments and any set of valid
fragments of this system can always be combined.  On the other hand, if the cluster
has a looped connectivity, where {\it A} and {\it D} may also bind, we would also have the
possibility to combine clusters {\it AB} and {\it D}, to produce {\it DAB}, {\it ABC} and {\it CD}.
This configuration is frustrated, as the fragments present cannot be combined to create any
targets without first breaking into smaller clusters.
\par
\begin{figure}[tb]
  \centering
  \includegraphics[width=75mm]{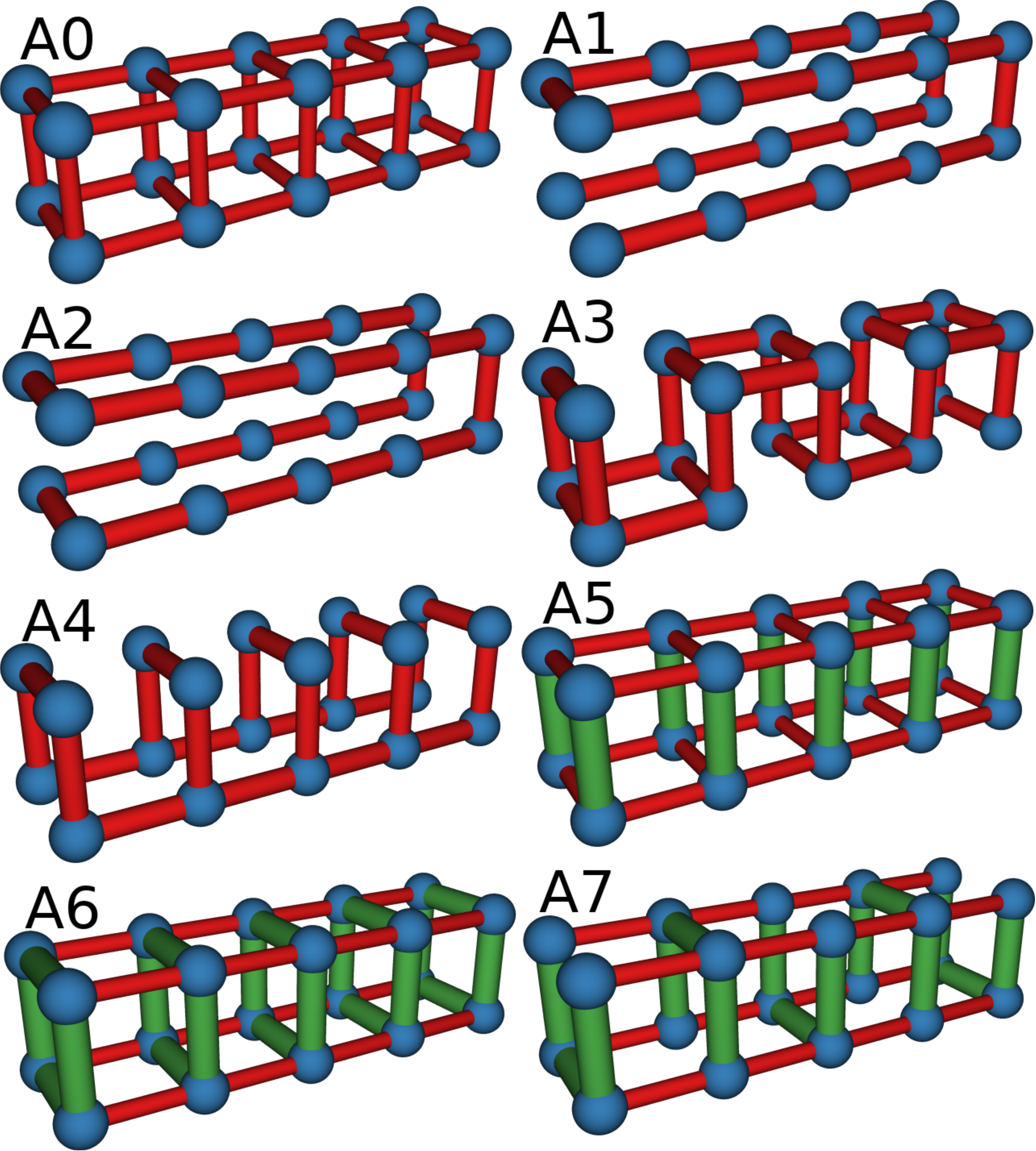}
  \caption{
    Bonding schemes for variations on target A simulated in this work.  A1
    and A2 are the linear and looped connectivities respectively.
    A3 contains loops of four particles and A4 contains loops of eight.
    Schemes A5--A7 contain a mixture of stronger (green) and weaker (red)
    bonds.  In A5, the strong bonds define disconnected dimers, while in
    A6 and A7 they define looped and linear tetramers, respectively.
  }
  \label{fig:compact_graphs}
\end{figure}
\begin{figure}[tb]
  \centering
  \includegraphics[width=75mm]{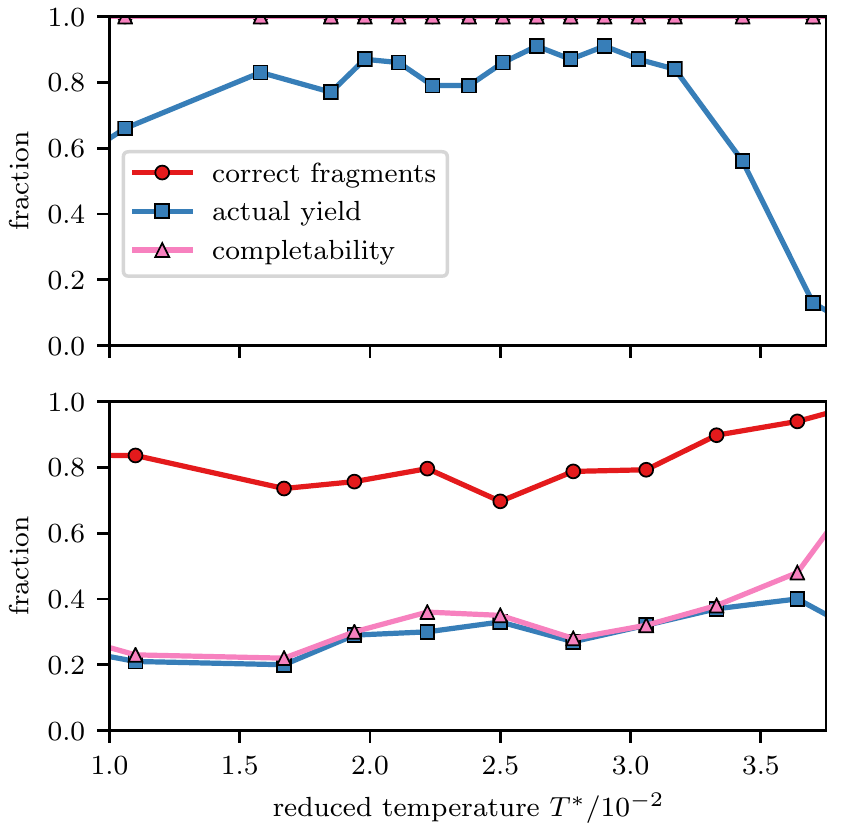}
  \caption{
    Yield, completability and fraction of correct fragments for the linearly
    connected A1 (top) and single-loop A2 (bottom) bonding schemes of target A.
  }
  \label{fig:frags_linear_oneloop}
\end{figure}

Two new connectivities of target A were designed, one linear, and one featuring
a single loop. These connectivities are shown in Fig.~\ref{fig:compact_graphs} A1
and A2, where blue spheres represent the location of particles, and red lines
the links between them. The yield plots for these connectivities are presented
in Fig.~\ref{fig:frags_linear_oneloop}.
For the linearly connected structure, the fraction of correct fragments and
the completability index remain at $1$ across the entire temperature range
simulated.  This confirms that fragments are assembling without error and that
there is never any incompatibility or frustration between fragments.
Nevertheless, it can take time for fragments to meet and bind.  At the end
of the assembly simulations, yield is uniformly high at about 80\% except
at high temperatures, where any system becomes thermodynamically
unstable with respect to dissociation.
Note that, due to removing many links, the temperatures at which the target
is stable are much lower than for the fully connected cluster A0
(Fig.~\ref{fig:frags_clusters}, top).  The shift is a potentially important consideration
in practice, since a self-assembled nano-structure or material may be required
to operate at a particular temperature in a given application.
\par
\begin{figure}[tb]
  \centering
  \includegraphics[width=75mm]{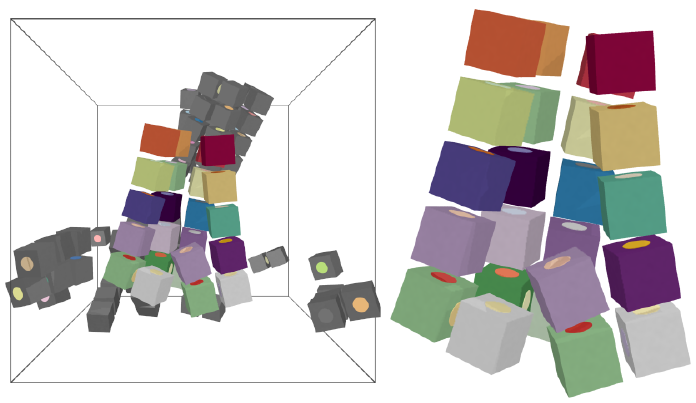}
  \caption{
    Left: A simulation snapshot of target A2 (with a single loop of bonds connecting
    particles in the target geometry).  An incorrect aggregate is highlighted in the
    center of the simulation box, where the flexibility of the structure has
    allowed the aggregate to grow rather than closing the loop and terminating
    assembly.  Right: An isolated view of the erroneous cluster.
  }
  \label{fig:snapshot_compact_oneloop}
\end{figure}

Since the linear bonding scheme contains the smallest number of bonds necessary
to hold the target together, a single break would split the cluster.  It is
therefore tempting at least to close the chain of bonds into a single loop,
which can be done by adding one bond to give scheme A2 in Fig.~\ref{fig:compact_graphs}.
The addition of this bond has a dramatic and detrimental effect on assembly,
as shown in the lower panel of Fig.~\ref{fig:frags_linear_oneloop}.  First, yield
at all temperatures closely follows the completability index, indicating that
further progress is limited by competition between incompatible fragments.
Secondly, the fraction of correct fragments has dropped to about $80\%$ across
most of the temperature range.  The incorrect fragments consist of chains where
individual bonds are approximately correct, but the structure has failed to
close up overall, leading to a spiral that continues to grow, as illustrated
in Fig.~\ref{fig:snapshot_compact_oneloop}.  This is also a consequence of the
loop in the bonding scheme, since growth is automatically
terminated at the ends of the chain in the linear bonding scheme.
\par
The likelihood of a structure like A2
failing to close depends on the flexibility of the individual links, which
determines the floppiness of the structure as a whole.
The formation of incorrect fragments therefore depends
on the details of the building-blocks and their bonding, and is a separate obstacle
to self-assembly from the competition between correct fragments.  Although the linear
bonding scheme A1 prevents run-away growth and leads to high yields, the final
structure is still floppy.  Hence, an entirely linear bonding scheme may still not
be satisfactory in practice, especially for larger structures.
\par
\begin{figure}[tbp]
  \centering
  \includegraphics[width=75mm]{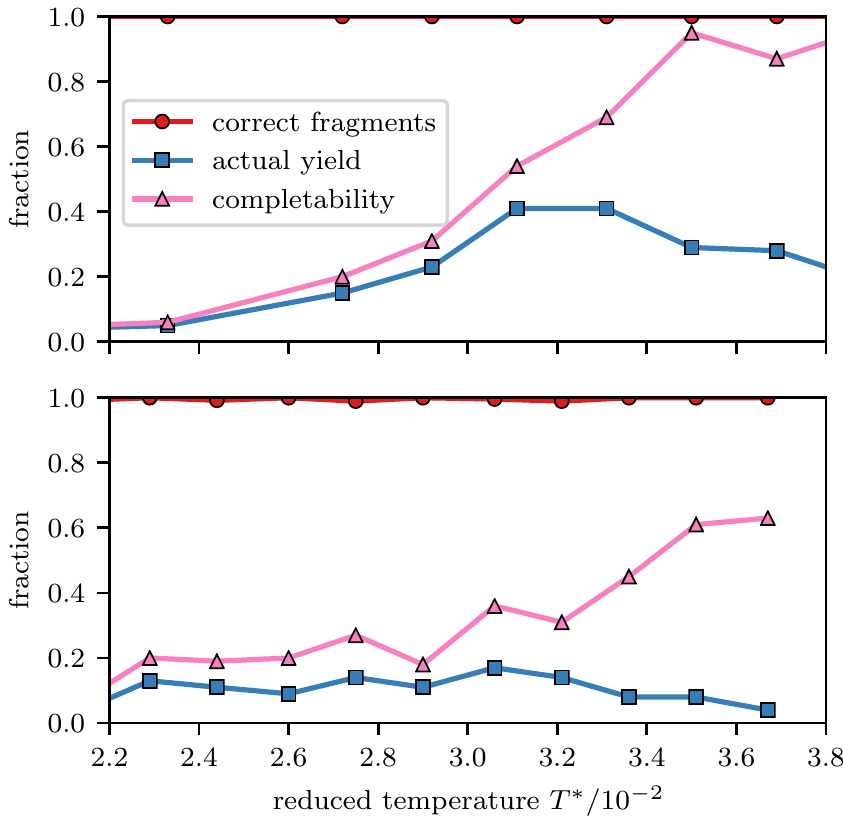}
  \caption{
    Yield, completability and fraction of correct fragments for bonding schemes
    A3 (top) and A4 (bottom) with intermediate connectivity (illustrated in
    Fig.~\ref{fig:compact_graphs}).
  }
  \label{fig:frags_rings}
\end{figure}

The fully connected bonding scheme A0 of target A effectively contains a very
large number of loops but assembles more effectively than the singly-looped
scheme A2.  Fig.~\ref{fig:compact_graphs} shows two possible intermediate
connectivities that lie between the extremes of A0 and A2.  Scheme A3 contains
multiple loops of four particles, while A4 has multiple loops of eight particles.
The results of self-assembly for A3 and A4 are shown in Fig.~\ref{fig:frags_rings}.
The smaller loops in both these schemes are both successful in eliminating the
incorrect structures that appeared for scheme A2.  The scheme with the smallest
loops, A3, is also successful at alleviating fragment competition compared to
A2.  However, neither of these intermediate schemes reaches peak yields as
high as the fully-connected scheme A0.  Any benefit from limiting the possible
combinations of competing fragments in the intermediate schemes is outweighed
by the fact that the particles only have a coordination number of 2 or 3
(compared to 3 or 4 in scheme A0).  The lower temperatures required for assembly
of A3 and A4 therefore intensify the competition between incompatible fragments.

\subsection{Heterogeneous bond energies}
\label{sec:str-res-hetero}
Altering the topology of the bonding network as in schemes A1--A4
amounts to switching edges on
and off in the graph of neighboring building-blocks so that $\varepsilon_{ij}$
takes one of only two values: $\varepsilon$ or $0$.  Finer control can
be exerted by altering the strength of individual bonds.  In DNA-based
systems, this is readily achieved through the choice of the nucleotides
or the length of the complementary sequences\cite{Sajfutdinow18a}.
\par
From previous work, we know that using different bond strengths can promote
hierarchical assembly pathways, since fragments where the monomers
are connected by strong links are
more likely to persist long enough to encounter and bind to each other than weakly
bound ones\cite{madge_design_2015}.  Having assembled to an intermediate stage,
the interactions between fragments are boosted if they can interact over multiple
sites in their larger combined interfaces, thereby driving the next stage of assembly.
Hierarchical assembly is normally envisaged
in the case where the structure itself is modular and
symmetrical\cite{groschel_guided_2013}.  However, such an approach is not
advantageous in all cases\cite{madge_design_2015,haxton_hierarchical_2013}.
Because of the uniqueness of building-blocks for fully addressable targets,
they have no formal modularity, but a hierarchical path could nevertheless
be used to limit the scope for fragment competition by controlling the fragments
that are likely to arise.
\par
This principle is demonstrated by the bonding network A5, illustrated in
Fig.~\ref{fig:compact_graphs}, where target A has been divided into 10
strongly-bound dimers that are linked to each other by weaker bonds.
There can be no competition in the pairing of building-blocks into
addressable dimers, so promoting early formation of these fragments
effectively reduces the number of unique components for assembly of the
full target from 20 to 10.
\par
\begin{figure}[tbp]
  \centering
  \includegraphics[width=75mm]{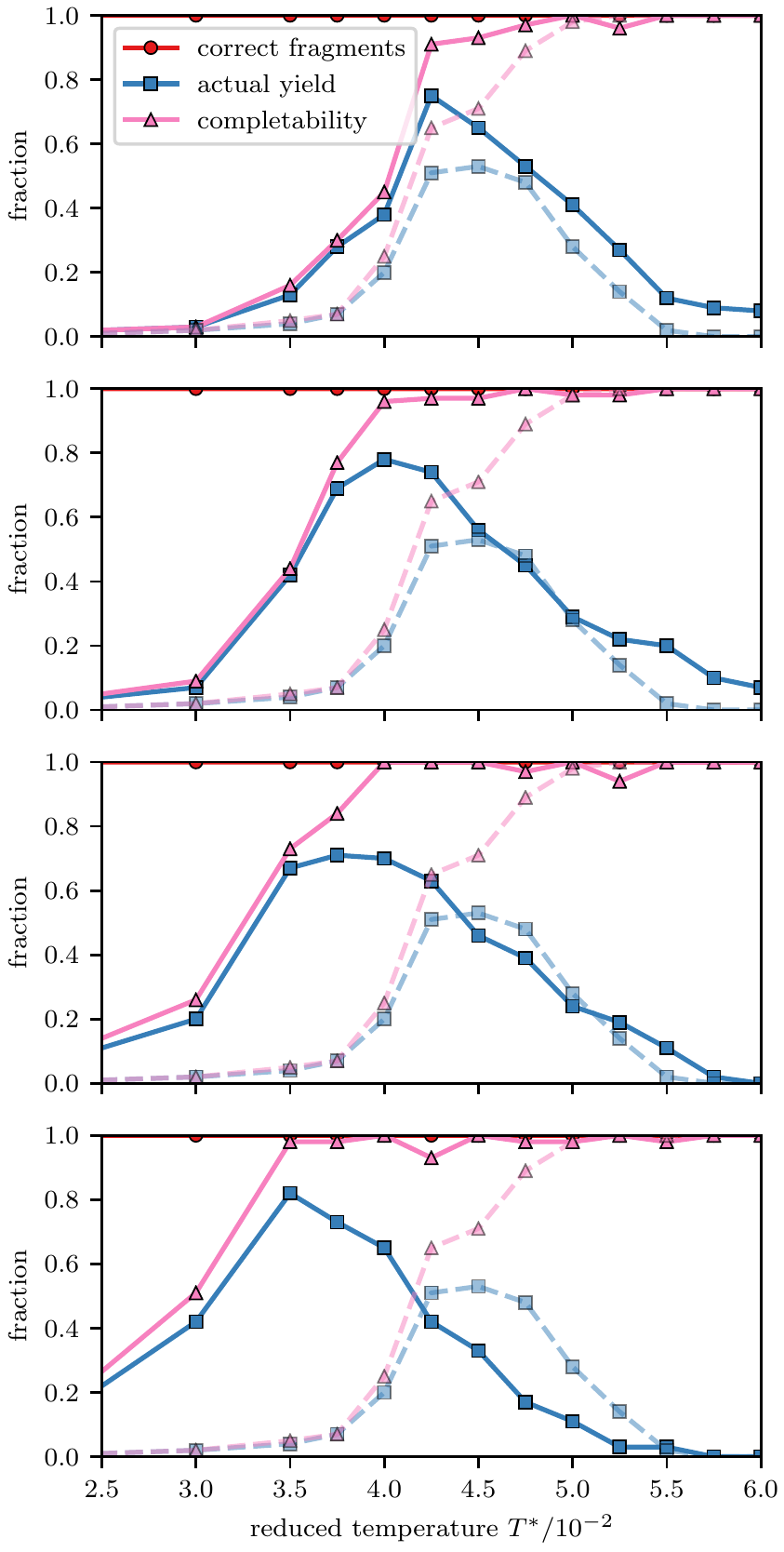}
  \caption{
    Yield, completability and fraction of correct fragments
    for bonding scheme A5, where target A has been
    divided into strongly-bound dimers.  From top to bottom the panels
    refer to strong:weak binding strength ratios of $1.5$, $2$, $2.5$ and $3$.
    The faint dashed lines show the results for bonds of equal strength
    (Fig.~\ref{fig:frags_clusters}).
  }
  \label{fig:frags_compact_dimers}
\end{figure}

The performance of model A5 is shown in Fig.~\ref{fig:frags_compact_dimers}
for a range of ratios of bond strengths (keeping the total binding energy
of the cluster fixed for comparison).  The results for the homogeneous
bond scheme A0 (Fig.~\ref{fig:frags_clusters}) are shown as faint lines for
comparison.  We see that dramatic improvement is possible with the inhomogeneous
bond scheme.  Fragment competition is indeed alleviated, as revealed by the
higher completability index at lower temperatures, especially for bonding-strength
ratios above $1.5$.  This has the effect of suppressing kinetic trapping
due to fragment competition, thereby broadening the temperature range for
successful assembly at the lower end and allowing higher peak yields (up to 80\%)
to be reached at the end of the chosen assembly time.  It is only at
strong:weak ratios of about $3$ that some of the high-temperature performance
starts to be lost, due to the weaker bonds then being insufficient to stabilize
the overall structure.
\par
\begin{figure*}
  \centering
  \includegraphics[width=130mm]{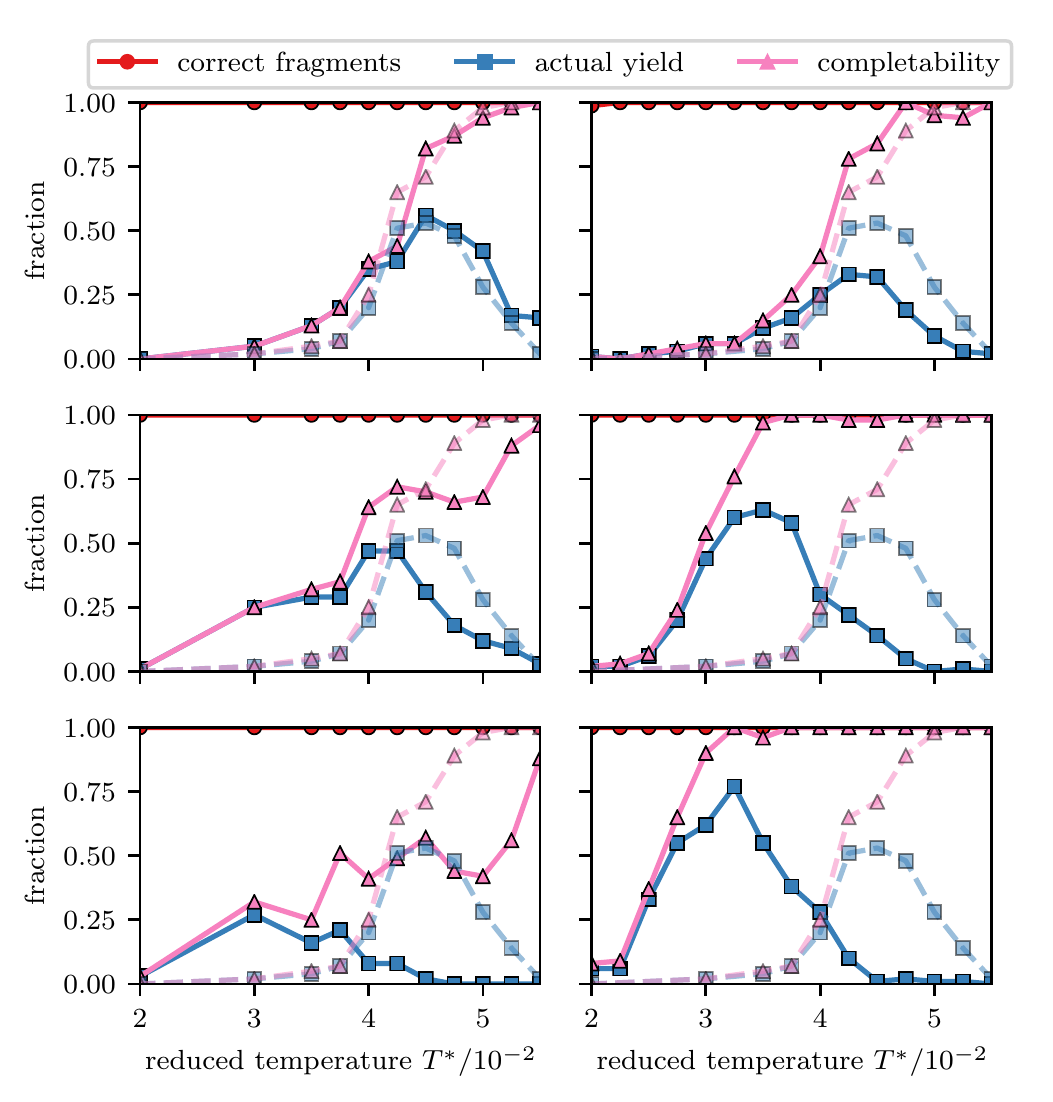}
  \caption{
    Yield, completability and fraction of correct fragments for bonding schemes
    A6 (left) and A7 (right), where target A has been divided into strongly-bound
    tetramers that are cyclically (A6) or linearly (A7) linked.
    From top
    to bottom, the panels refer to strong:weak binding strength ratios of
    $1.5$, $2.5$ and $3$.  
    The faint dashed lines show the results for bonds of equal strength
    (Fig.~\ref{fig:frags_clusters}).
  }
  \label{fig:frags_compact_tetramers}
\end{figure*}

Another possible hierarchical scheme for target A would be to divide it into
five slices, each of which is a square tetramer defined by strong bonds,
linked to each other by weaker bonds.  The tetramers may be cyclically
connected, as in scheme A6 of Fig.~\ref{fig:compact_graphs}, or linearly
connected (omitting one edge of each square), as in scheme A7.  The
performance of both schemes is shown in Fig.~\ref{fig:frags_compact_tetramers},
again comparing with the heterogeneous scheme A0 from Fig.~\ref{fig:frags_clusters}.
\par
The cyclic tetramer scheme A6 has modest success in broadening the temperature
range of assembly for bond-strength ratios around $2$--$2.5$, but peak yield
is not improved.  The scheme suffers from fragment competition in the
first stage of assembly, due to the cycle of bonds in the tetramers, thereby
stalling the second stage.  The linearly bonded tetramer scheme A7 is much
more effective, since fragment competition is avoided at both stages of assembly,
provided that the stages are energetically well separated.  At sufficiently low
temperature, fragment competition does return because even fragments bound by the
weaker links then become long-lived and impede formation of the tetramers.  Overall,
peak yields approach those of the dimer scheme A5, but with a slightly narrower
range of assembly temperatures.
\par
\begin{figure}[tb]
  \centering
  \includegraphics[width=60mm]{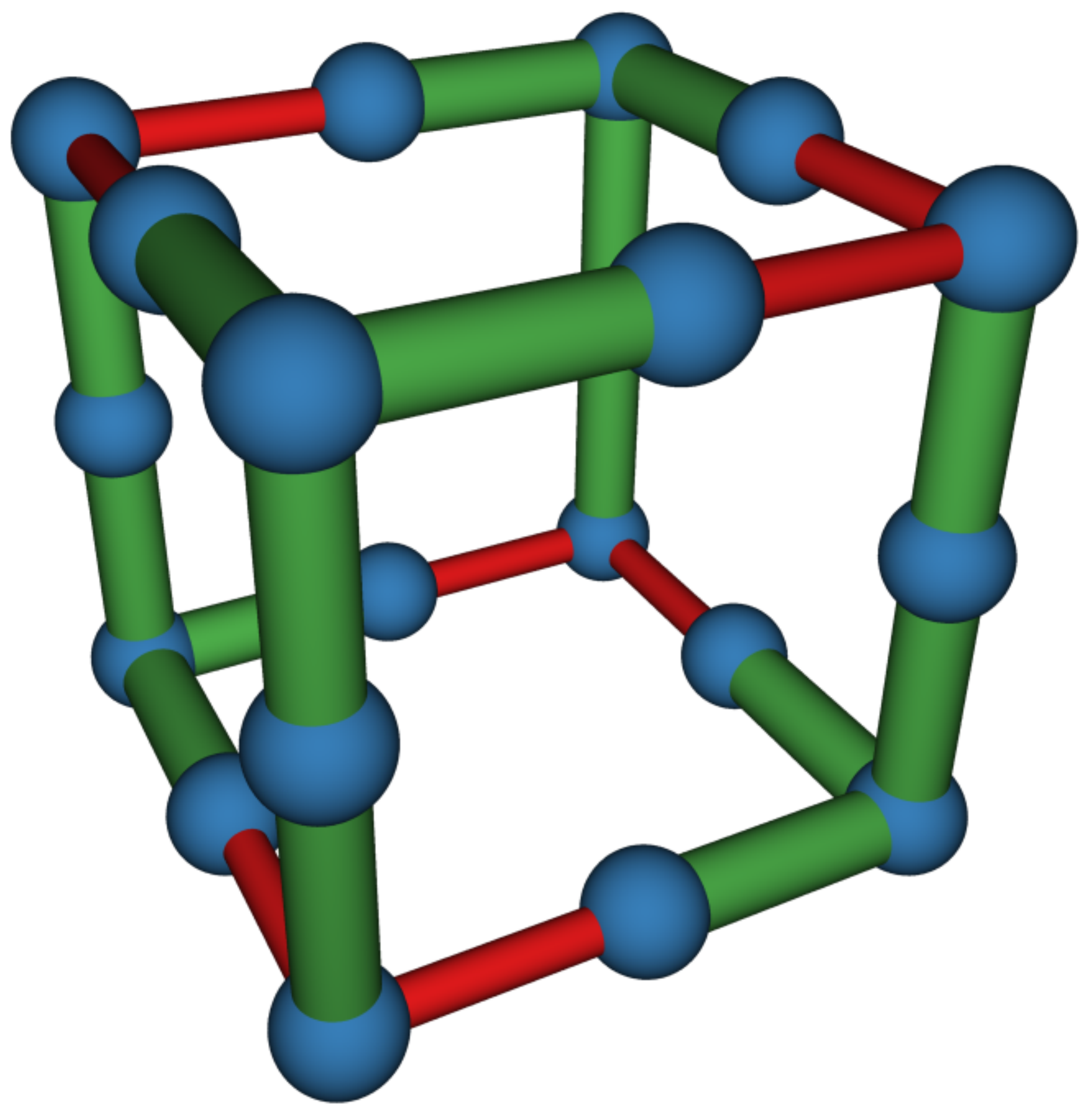}
  \caption{
    Heterogeneous bonding scheme B1 for target B, dividing the structure into
    four pentamers.  Green and red
    cylinders indicate strong and weak links, respectively.
  }
  \label{fig:open_strong}
\end{figure}
\begin{figure}[tbp]
  \centering
  \includegraphics[width=75mm]{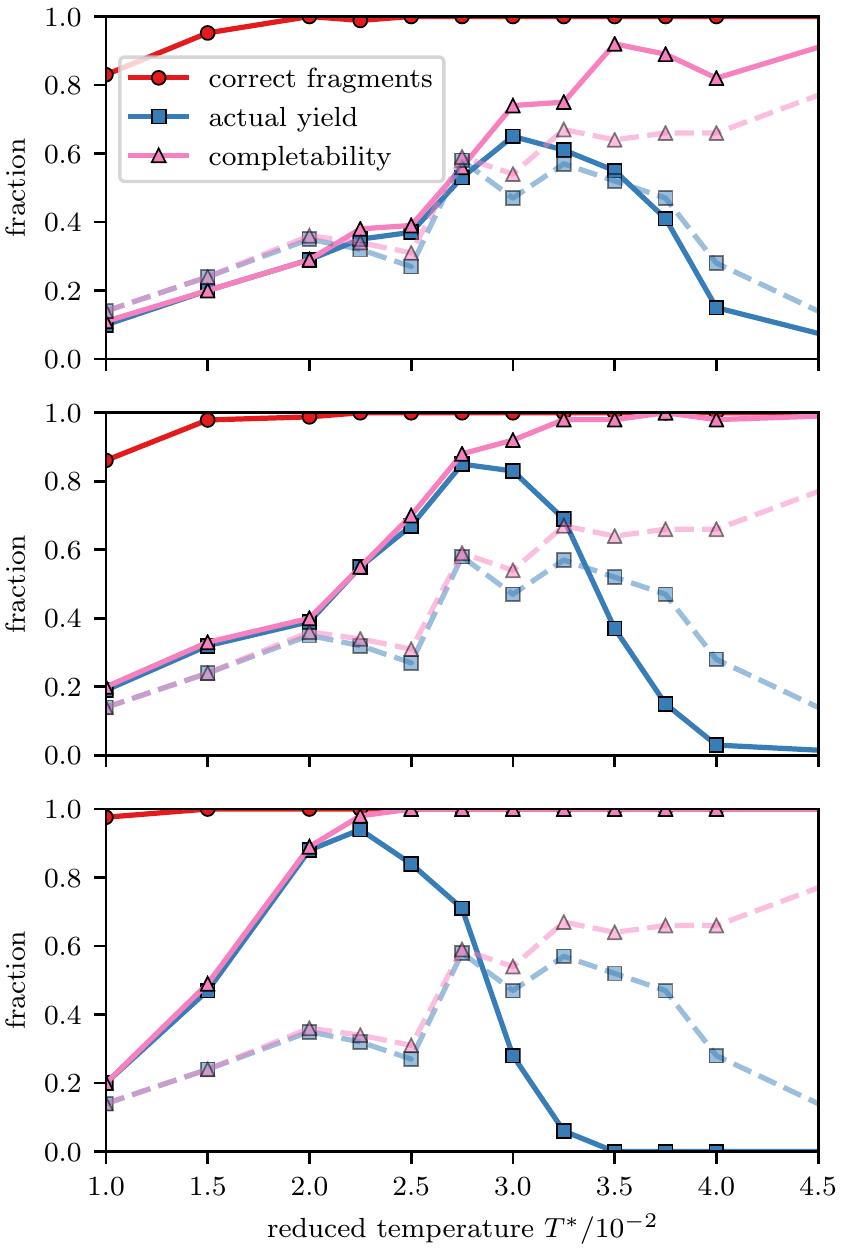}
  \caption{
    Yield, completability and fraction of correct fragments for the heterogeneous
    bonding scheme of target B1 illustrated in Fig.~\ref{fig:open_strong}.
    From top
    to bottom, the panels refer to strong:weak binding strength ratios of
    $1.5$, $2$ and $3$.  
    The faint dashed lines the results for the homogeneous bond scheme
    (Fig.~\ref{fig:frags_clusters}).
  }
  \label{fig:frags_open_pentamers}
\end{figure}

A hierarchical scheme can also be used to improve the self-assembly of the more
open target B.  On the basis of the results so far, we expect the most successful
scheme to omit bonding loops at least in the first stage of assembly.  A
symmetrical scheme of heterogeneous bonding is shown in Fig.~\ref{fig:open_strong}.
The graph of bonds in the strongly-linked pentamer fragments is now branched but
still non-cyclic, thereby avoiding the possibility of fragment competition
in formation of the first-stage fragments.
Yield curves for this scheme are presented in
Fig.~\ref{fig:frags_open_pentamers}.  As predicted, the hierarchical scheme with
non-cyclic fragments suffers less from fragment competition at low temperatures
than the homogeneously bonded scheme.  The peak yield is over $90\%$ at the
optimum temperature for a bond-strength ratio of $3$.  For this target,
assembly temperatures are shifted noticeably downwards in the heterogeneous
scheme because the structure must still be held together by a relatively small
number of weak links.

\subsection{Randomized schemes}

The placements of strong and weak links in schemes A5, A7 and B1 were guided
by our observation that linear connectivities avoid competition between
fragments and that competition is particularly undesirable in the early
stages of assembly.
To test whether this intuitive approach is strictly necessary, we have
briefly examined schemes where a given number of strong bonds are distributed
randomly in a fully connected network of bonds for target A.  The procedure
is described in more detail in the Supporting Information and the results are
shown in Fig.~S1.  The main effect of the randomly placed strong bonds
is to shift the assembly temperature downwards slightly.  For a
small-to-intermediate number of strong bonds, there can be a
small increase in peak yield and a slight widening of the temperature range
for successful assembly.  However, the full advantages of a strategically chosen
network are not realized by random placement.
\par
Bond strengths do not have to be restricted to two values and, in some applications
such as DNA bricks, naturally have a continuous distribution about their mean.
Theoretical and computational work by Jacobs {\it et al.}\cite{jacobs_rational_2015}
has predicted that a spread of bond energies can assist addressable assembly by
stabilizing small, floppy structures and lowering the nucleation barrier for assembly.
We have tested Gaussian distributions of bond strengths with a selection of
different widths for target A, as described in more detail in the supporting
material.  For this modest-sized target structure and the choice of assembly
protocol used in the present work, we found little effect of random bond-strength
heterogeneity (Fig.~S2).  Any slight alleviation of fragment competition was
accompanied by a slight loss of peak yield.

\section{Conclusions}
\label{sec:str-conclusion}

The aim of any self-assembly process is to produce the maximum number of
complete, defect-free target structures in a reasonable time.
In this work, we have examined one
of the ubiquitous potential obstacles to this goal, which is the formation
of, and frustration between
multiple incomplete fragments.  This possibility is particularly pertinent
in the context of addressable assembly, where only specific combinations of
building-blocks are permitted.  In contrast to the majority of computational
studies of addressable assembly, we have explicitly included competition between
multiple copies of the target structure in our simulations.
\par
We have quantified the incompatibility of fragments at a given stage of assembly
by introducing a completability index, which determines the globally optimal way of
combining fragments.  It is important to realize that this index is not a static
quantity and is likely to change nonmonotonically over the course of self-assembly.
If assembly is initialized from disconnected
monomers, the completability begins at unity.  Similarly, if all components
are successfully incorporated into defect-free targets then the completability
ends at unity.  However, the completability index drops when it detects
intermediate fragments that cannot be combined in principle, due to their
incompatible compositions.  This competition may slow down assembly because
some fragments must break up before progress can be made.  Depending on the
severity of the competition, it may also limit the final yield of the target
that can be achieved in practice, even if no erroneous structures have formed.
In this work, we have compared all schemes after a fixed assembly time.  This
provides a way to judge the efficiency of assembly, even though different
schemes may be advancing at somewhat different rates at any given cut-off
time\cite{madge_design_2015}.
\par
An important general result that the completability index makes clear is
that fragment competition is ruled out in any system where the network of
bonds contains no cycles.  In such cases, correctly formed fragments can never
be incompatible with completion of the maximum number of targets due to their
composition alone.  In practice, other considerations also come into
play.  Depending on the nature of the binding, cycle-free designs are likely
to be too floppy.  Except at very low temperature, they are also vulnerable
to dissociation because every edge is a cut-edge in the graph of bonds;
hence, disruption of any bond breaks the structure.
Floppiness can also lead to incorrect fragments due to uncontrolled growth.
In the present work, such growth is one of the few causes of invalid aggregation
because we have taken the selectivity of building-block interactions to
be perfect.  In the presence of cross-interactions, aggregation is an
additional obstacle to self-assembly\cite{madge_minimal_2017}.
\par
Building on these observations, we have shown that a successful strategy for
efficient assembly is to encourage partially hierarchical assembly pathways
in which the first steps involve formation
of cycle-free fragments.  If the stages of assembly can be cleanly separated
then this approach effectively reduces the complexity of the later stages to
assembly of a smaller number of addressable fragments.  As previous
work\cite{madge_design_2015} and the contrasting targets A and B in this
contribution show, the best choice of stepwise path depends on the specific
target, including steric considerations.
We note that very recent experimental developments now make it possible to
follow the assembly of addressable structures of the size we have considered
here and to manipulate pathways by selectively altering bond
strengths\cite{Sajfutdinow18a}.
\par
In this work, the assembly protocol has been an instantaneous quench from
a low-density dispersion of monomers at high temperature to a range of fixed
assembly temperatures.  This has allowed us to judge the success of
self-assembly not only by the peak yield after a given time, but also by the
range of temperatures over which assembly is successful even if sub-optimal.
Robustness with respect to the precise conditions is a desirable property of 
a practical self-assembling system, since performance is then less
dependent on fine-tuning.
Addressable assembly of DNA bricks can indeed operate successfully by
incubation at fixed temperature\cite{ke_dna_2014}, although a gradually
decreasing temperature ramp\cite{Ong17a} can have the advantage of encouraging
nucleation at a small number of sites at high temperature, followed
by growth to completion at a lower
temperature\cite{jacobs_rational_2015,jacobs_communication:_2015}.
We note that a time-dependent protocol could be used to enhance the
partly hierarchical approach suggested here, since it would allow a greater
separation of energy scales to come into play in succession, thereby
avoiding interference between the different stages of assembly.  With such
an approach, the results of this work could be used to control fragment competition
at each scale in the proposed ensemble of pathways.

%%%%%%%%%%%%%%%%%%%%%%%%%%%%%%%%%%%%%%%%%%%%%%%%%%%%%%%%%%%%%%%%%%%%%
%% The same is true for Supporting Information, which should use the
%% suppinfo environment.
%%%%%%%%%%%%%%%%%%%%%%%%%%%%%%%%%%%%%%%%%%%%%%%%%%%%%%%%%%%%%%%%%%%%%
\begin{suppinfo}

Self-assembly results for randomly placed strong bonds (Figure S1);
Self-assembly results for a Gaussian distribution of bond energies
(Figure S2); worked example of completability algorithm.

\end{suppinfo}

%%%%%%%%%%%%%%%%%%%%%%%%%%%%%%%%%%%%%%%%%%%%%%%%%%%%%%%%%%%%%%%%%%%%%
%% The appropriate \bibliography command should be placed here.
%% Notice that the class file automatically sets \bibliographystyle
%% and also names the section correctly.
%%%%%%%%%%%%%%%%%%%%%%%%%%%%%%%%%%%%%%%%%%%%%%%%%%%%%%%%%%%%%%%%%%%%%
% \bibliography{bibliography}

\providecommand{\latin}[1]{#1}
\makeatletter
\providecommand{\doi}
  {\begingroup\let\do\@makeother\dospecials
  \catcode`\{=1 \catcode`\}=2 \doi@aux}
\providecommand{\doi@aux}[1]{\endgroup\texttt{#1}}
\makeatother
\providecommand*\mcitethebibliography{\thebibliography}
\csname @ifundefined\endcsname{endmcitethebibliography}
  {\let\endmcitethebibliography\endthebibliography}{}

\end{document}